\documentclass[conference]{IEEEtran}
\IEEEoverridecommandlockouts
\usepackage{amsmath,amssymb,amsfonts}
\usepackage{algorithmic}
\usepackage{graphicx}
\usepackage{textcomp}
\usepackage{xcolor}

\usepackage{float}

\usepackage[hidelinks]{hyperref}
\usepackage{color}
\usepackage{orcidlink}

\usepackage[style=ieee, maxbibnames=6]{biblatex}
\def\BibTeX{{\rm B\kern-.05em{\sc i\kern-.025em b}\kern-.08em
    T\kern-.1667em\lower.7ex\hbox{E}\kern-.125emX}}
    
\begin{document}

\title{Serverless Approach to Running Resource-Intensive STAR Aligner}

\author{\IEEEauthorblockN
{
Piotr Kica\IEEEauthorrefmark{1}\IEEEauthorrefmark{2}\orcidlink{0009-0001-9269-6804},
Michał Orzechowski\IEEEauthorrefmark{1}\IEEEauthorrefmark{2}\IEEEauthorrefmark{3}\orcidlink{0000-0002-8558-1283},
Maciej Malawski\IEEEauthorrefmark{1}\IEEEauthorrefmark{2}\orcidlink{0000-0001-6005-0243}
}
\IEEEauthorblockA{\IEEEauthorrefmark{1}Sano Centre for Computational Medicine, Krak\'ow, Poland\\}
\IEEEauthorblockA{\IEEEauthorrefmark{2}Faculty of Computer Science, AGH University of Krak\'ow, Poland\\}
\IEEEauthorblockA{\IEEEauthorrefmark{3}Academic Computer Centre Cyfronet AGH, Krak\'ow, Poland\\}
}

\maketitle

\begin{abstract}
The application of serverless computing for alignment of RNA-sequences can improve many existing bioinformatics workflows by reducing operational costs and execution times. This work analyzes the applicability of serverless services for running the STAR aligner, which is known for its accuracy and large memory requirement. This presents a challenge, as serverless services were designed for light and short tasks. Nevertheless, we successfully deploy a STAR-based pipeline on AWS ECS service, propose multiple optimizations, and perform experiment with 17 TBs of data. Results are compared against standard virtual machine (VM) based solution showing that serverless is a valid alternative for small-scale batch processing. However, in large-scale where efficiency matters the most, VMs are still recommended. \\
\end{abstract}

\begin{IEEEkeywords}
Serverless, Alignment, Transcriptomics, Cloud, STAR, AWS, Container-as-a-service
\end{IEEEkeywords}

\section{Introduction}
Alignment of RNA sequences to a reference genome is crucial for understanding gene expression, identifying genetic variants, and performing transcriptome studies. One of the best aligners used for this task is STAR~\cite{dobin2013star} - a resource-intensive software and has been widely used in the transcriptomics domain because of its speed and accuracy. 

Serverless services, including Function-as-a-Service (FaaS) and Container-as-a-Service (CaaS) execution models, have been established as a suitable solution for running multiple types of tasks, including batch processing and highly dynamic workloads~\cite{malawski-serverless}. The advantages over more traditional services come from the high scalability, lower operational complexity, and the pay-as-you-use pricing model. However, such services come with different limits and some may be more suitable than others for processing RNA-sequences with STAR aligner. 

Successful attempts have been made to run other aligners on serverless computing services. The authors of~\cite{cinaglia2022serverless,cinaglia2023massive} run a much less memory-intensive aligner -- HiSat2 -- using AWS Lambda. Moreover, by splitting the input FASTQ files into chunks, they parallelize the computation and significantly speed up the processing of a single file. However, STAR is still considered to be more accurate~\cite{baruzzo2017simulation,bianchi2023comparing} and can have better processing time~\cite{bianchi2023comparing}.

The STAR aligner poses a significant challenge in running efficiently in serverless mode. The main difficulty is that STAR requires a precomputed index built on a reference genome which has to be loaded into system's memory. In the case of the human genome, such index size is about 30GB which exceeds many services' limits. 

The research goals are as follows:
\begin{itemize}
    \item Find suitable serverless services for STAR alignment,
    \item Perform experiment with serverless STAR aligner,
    \item Compare the cost of the serverless solution with more traditional approaches,
    \item Identify possible optimizations,
    \item Describe use cases for serverless STAR deployment.
\end{itemize}

\section{Challenges addressed and solution architecture}

\subsection{Pipeline description and resource requirements}
\begin{figure*}[tb]
    \centering
    \includegraphics[width=\linewidth]{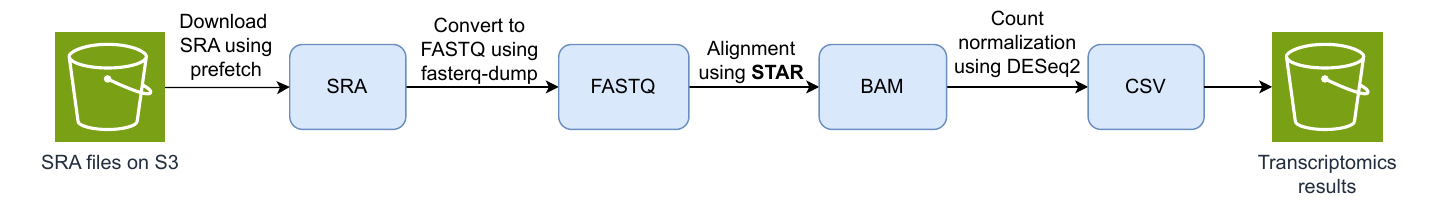}
    \caption{STAR-based transcriptomics pipeline. Processing of public RNA-seq data from NCBI.}
    \label{fig:TAtlas_pipeline}
\end{figure*}

We use a STAR-based pipeline described in~\cite{kica2024optimizing} and presented in Fig.~\ref{fig:TAtlas_pipeline}. It allows accessing public RNA-sequences from NCBI database with commonly used SRA-Toolkit software. The first step is to download the compressed file in \textit{SRA} format. Then it is converted into FASTQ format which is required for STAR alignment. FASTQ files can be very large, up to 220 GB for the experiment in Section~\ref{sec:experiment}. Moreover, the \texttt{fasterq-dump} tool requires additional storage during SRA to FASTQ conversion. 

\begin{table*}[!htb]
\caption{Applicability of leading serverless services for RNA-Seq alignment with STAR.}
\resizebox{\textwidth}{!}{%
    \centering
    \begin{tabular}{|c|c|c|c|c|}
    \hline
    Service & Max execution time & Max RAM & Available storage & Can it run STAR? \\ \hline
    AWS ECS + Fargate & 14 days & 120 GB & \begin{tabular}[c]{@{}c@{}} Block storage {[}\textgreater{}TBs{]} or \\ Ephemeral storage {[}\textless{}200GiB{]}\end{tabular} & Yes \\ \hline
    AWS Lambda & 15 min & 10 GB & NFS or object storage & No \\ \hline
    Google Cloud Run & 1h / 7 days (preview) & 32 GiB & NFS or object storage & \begin{tabular}[c]{@{}c@{}} Only with small genomes \\ and small FASTQ files\end{tabular} \\ \hline
    Azure Functions & 10 min & 1.5 GB & NFS or object storage & No \\ \hline
    \end{tabular}
    \label{tab:services_analysis}
}  
\end{table*}

The most important step, which is the alignment with STAR, usually constitutes about 70-75\% of the total pipeline execution time. In this step STAR uses an index data structure which is generated beforehand on a reference genome. Its size depends on the genome origin, e.g. an index built on a human genome is 30GB in size. Since the index must be fully loaded into the system's memory, this poses a challenge for efficient computations. Once loaded, it can be reused for future alignments. In the case of a human genome, it takes about 5-10 minutes to load the index. The pipeline code is publicly available on GitHub under the MIT license~\cite{neardata_repo}. 

\subsection{Serverless services comparison}
In Table~\ref{tab:services_analysis} we compare the most commonly used serverless computing services in the context of running STAR aligner. Lightweight solutions such as AWS Lambda or Azure Functions are ruled out due to unmet RAM requirments. Moreover, too-short maximum execution times are highly limiting since they may not be enough for processing a single FASTQ file. Furthermore, none of those services has enough ephemeral storage or free RAM for FASTQ files. However commonly practiced, using object storage is not suitable for serverless high-performance workloads due to slow I/O times~\cite{arjona2023scaling}. Also, running multiple alignments favors long-running services as it reduces initialization cost associated with loading the index.

Therefore, the best solution is to use AWS Elastic Container Service (ECS) in Fargate mode as it fulfills all the technical requirements for running serverless STAR alignment.

\subsection{Serverless deployment on ECS Fargate}
\begin{figure}[tb]
    \centering
    \includegraphics[width=\linewidth]{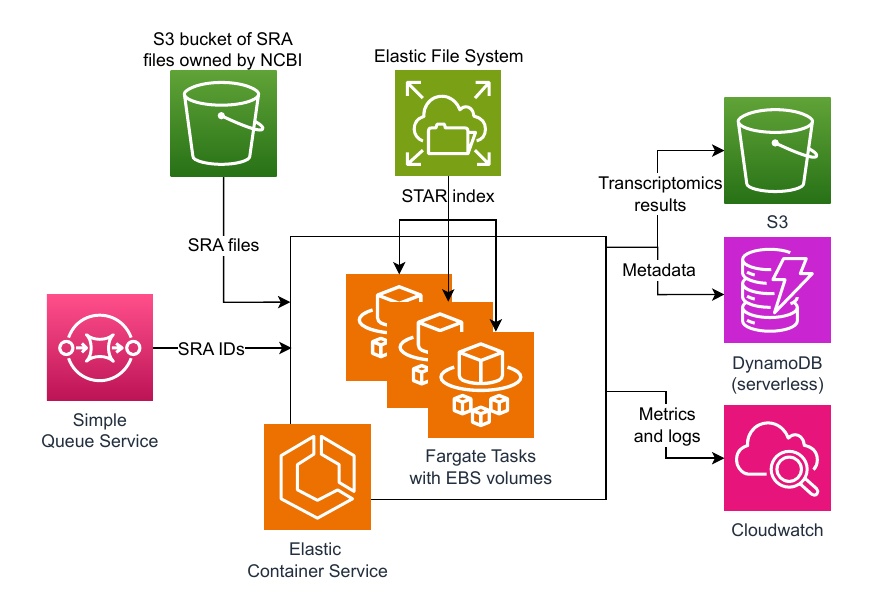}
    \caption{Cloud architecture for serverless STAR alignment on AWS.}
    \label{fig:ECS_fargate_architecture}
\end{figure}

In Fig.~\ref{fig:ECS_fargate_architecture} we present the architecture designed for batch processing RNA-sequences using ECS in Fargate mode. The STAR-based pipeline (Fig.~\ref{fig:TAtlas_pipeline}) is embedded in the container image used for ECS tasks. Each task initially mounts the Elastic File System volume that contains the precomputed index and loads it into memory to be used for STAR alignment. By using EBS volumes, each task has enough block storage for processing FASTQ files without I/O bottleneck.

\section{Experiment}
\label{sec:experiment}
In order to validate the solution and compare it with the traditional VM-based approach, we perform two experiments in which we process the same data set with a similar configuration but using different compute services - ECS in Fargate mode and EC2 (Elastic Compute Cloud). 

Experiment configuration:
\begin{itemize}
    \item ECS: 20 task instances in us-east-1
    \item ECS Task resource: 8 vCPU, 48 GB of RAM
    \item EC2 instance type: 8 vCPU, 64 GB of RAM (r7a.2xlarge)
    \item EBS: 550GB, GP3, 500MiB/s, 3000IOPS
    \item Input: 1000 \textit{SRA} files (2.35 TB total size, max=28.7GB)
    \item Index: Generated on Toplevel human genome, Release 111, 29.5GB size.
\end{itemize}

\subsection{Results}
The experiment took 207h on ECS and 138.6h on EC2. The difference is caused by slower STAR alignment on ECS tasks. The ECS subpar performance is caused by the usage of older generation CPU models. The ECS tasks in the experiment used processors such as AMD EPYC 7R13 / Intel Xeon 8259CL / Intel Xeon 8175M which are found in 5th and 6th generation of EC2 instances. Serverless services do not provide control over the CPU models allocated for execution.
For EC2 experiment instance type from 7th generation was selected, which has one of the fastest CPU models available (AMD EPYC 9R14). In this case, faster processing contributes to further savings on the usage time of the to attached EBS volumes.

Compared to EC2, the ECS simplifies the allocation of sufficient RAM memory for a task by allowing for more precise resource specification, thereby reducing the risk of overprovisioning. Therefore, we use 48 GB instead of 64 GB as in the case of the EC2 experiment. Interestingly, the result was the unsuccessful alignment of 11 FASTQ files (out of 1000) due to insufficient memory.

The total estimated cost to process the data set using ECS in Fargate mode was 127\$ including 15\$ for EBS volumes. The estimated cost for processing with EC2 instances was 96\$ including 11\$ for EBS volumes. In total 17.3TB of FASTQ data have been processed with STAR.

\subsection{Possible optimizations}
Similarly to the EC2 service, one of the possible cost optimizations that can be used with ECS is Fargate Spot. In this mode, we get ephemeral instances that can be terminated with short notice. In turn, we receive a discount on computing resources that can be up to 70\% cheaper. Estimates show that the cost of the experiment can be reduced to about 50\$. 

As described in~\cite{kica2024optimizing} it is possible to leverage intermediate metrics acquired during alignment to terminate the processing of low-quality sequences, thus saving computational resources. The authors report a 23\% reduction in the total STAR alignment time.

\section{Conclusions}
The developed architecture shows that running STAR aligner on serverless services is possible, however, only the ECS Fargate service is suitable for this resource-intensive workload. Other services do not allow for sufficient memory, or storage for FASTQ processing, and are limited by maximum execution time. 

The experiment showed that this approach is valid but more expensive than the EC2-based solution. Therefore, it should be used for small-/medium-size datasets where serverless benefits such as lower operational costs and a pay-as-you-use pricing model are crucial. Furthermore, self-hosted serverless or workflow-related solutions such as Argo Workflows, Knative or OpenFaaS involve additional operational effort for manual setup due to operational overhead.

Promising results could be achieved by applying the divide-and-conquer strategy, where a FASTQ file is split into chunks and processed in parallel using serverless containers. For this purpose, we can leverage existing tools such as Dataplug that facilitate efficient partitioning of scientific data formats, including FASTQ~\cite{arjona2024dataplug}. With this solution, we could significantly reduce the alignment time of a single sequence. 

\section*{Acknowledgment}
The publication is supported by the Polish Minister of Science and Higher Education, contract number MEiN/2023/DIR/3796;  EU Horizon 2020 Teaming grant agreement No 857533; IRAP program of the Foundation for Polish Science; and EU Horizon Europe grant NEARDATA No 101092644.

\printbibliography

\end{document}